\documentstyle[preprint,aps]{revtex}
\title{Enhancement of low-${\bbox m_t}$ kaons in AGS heavy-ion collisions}
\author{G. Q. Li and  C. M. Ko}
\address{Cyclotron Institute and Physics Department\\
Texas A\&M University, College Station, Texas 77843}
\begin{document}
\maketitle

\begin{abstract}
In the relativistic transport model, we show that the recently observed
enhancement of low-$m_t$ kaons ($K^+$ and $K^-$) in Si+Pb collisions
at AGS can be explained if a density isomer is introduced
in the nuclear equation-of-state.
\end{abstract}

\section{Introduction}
An interesting experimental observation in relativistic heavy-ion
collisions at AGS energies is the enhancement of low-$m_t$
kaons (both $K^+$ and $K^-$) in Si+Pb collisions at 14.6 (GeV/c)/nucleon
\cite{STA94}. The spectra of these extremely `cold' (in terms of
their transverse mass $m_t$) kaons can be characterized by an inverse
slope parameter (temperature)
as low as 15 MeV, which is about one order-of-magnitude smaller
than the temperature of `normal' kaons measured in earlier AGS
experiments \cite{ABB90}.  These low $m_t$ kaons cannot be
obtained in conventional calculations such as the relativistic quantum
molecular dynamics \cite{MATT89}.
At lower incident energies, a similar
enhancement of low-$p_t$ pions has already been observed at
Bevalac \cite{LBL} and SIS energies \cite{GSI}, and one of
the plausible explanations for this enhancement, as
put forward by Xiong {\it et al.} \cite{XIONG93}, is
based on the attractive pion potential due to the $\Delta$-hole
polarization.

Very recently, Koch \cite{KOCH94} has tried to explain the enhancement
of low-$m_t$ kaons at AGS energies by a similar mechanism
as in Ref. \cite{XIONG93}, i.e., the attractive kaon potential.
When a kaon `climbs' out of a potential well, it has to lose its
kinetic energy and thus gets cooled down.  It was found in Ref. \cite{KOCH94},
however, that in order for the attractive potential to be effective
in cooling down the kaons, the baryon fireball, which is the source for
the kaon potential, has to expand slower than kaons,
so that the latter will have the chance to climb out the potential well
before it disappears.

In a relativistic transport model \cite{KO87,MOS88} based on
the $\sigma$-$\omega$ model \cite{QHD}
for the nucleon-nucleon interaction, as the one used
in this work and in Ref. \cite{KOCH94}, the expansion
of the fireball is controlled by the delicate balance (or rather unbalance)
between the scalar ($\sigma$) attraction  and vector ($\omega$) and thermal
repulsion.
In usual relativistic transport model, the scalar and vector coupling
constants,
which are determined by fitting the properties (saturation density,
binding energy, effective mass and incompressibility) of normal nuclear matter,
are assumed to be density and temperature independent. For a fireball
at about six times normal nuclear matter density and
a temperature of about 170 MeV,
as expected for the AGS energies \cite{MATT89,PANG92},
the scalar attraction is overwhelmed
by the vector and thermal repulsion, as the former tends to saturate
at high density due to relativistic effects. The fireball would thus
expand rather fast and kaons cannot be cooled down effectively.

To slow down the expansion of the fireball, Koch \cite{KOCH94} has
suggested that the system initially goes through the chiral restoration
transition, which leads to a softening of the equation-of-state and thus
slows down the expansion of the fireball. This has been achieved in
Ref. \cite{KOCH94} by reducing the vector coupling constant to about
half of the value used in the original
Walecka model \cite{QHD}. Using the coupling constants given in Ref.
\cite{KOCH94}, however, the nuclear matter would have saturated around
0.6 fm$^{-3}$ with a binding energy of about 250 MeV, which is clearly
unrealistic.

In this paper, we propose that the slowing down of the fireball expansion
is also possible if there is a density isomer
at high densities and if the fireball is initially at this abnormal state.
Lee and Wick \cite{LEE74} observed that the chiral sigma model can
lead to an abnormal state at high densities. The binding energy of this state
can be very large, leading to a secondary minimum in the nuclear matter
equation-of-state. The chiral sigma model, however, cannot describe
correctly the saturation properties of the nuclear matter. In a relativistic
mean-field theory that includes the $\Delta$ degree of freedom, Boguta
\cite{BOG82} has shown that it is possible to $simultaneously$ describe
the nuclear matter properties at saturation density and predict the
existence of a density isomer at high density. This has been realised
in Ref. \cite{BOG82} by assuming that the scalar interaction of the $\Delta$
is stronger than that of the nucleon. In the present exploratory study, we will
simulate the effect of the density isomer by introducing a density-dependent
vector coupling constant in the non-linear $\sigma$-$\omega$
model, which gives correct saturation properties of the nuclear matter
at $\rho _0=0.16$ fm$^{-3}$ and leads to
a secondary minimum around (5-6)$\rho _0$. More consistently, one
would like to combine the chiral sigma model with the Walecka model, so
that both the nuclear matter saturation and the density isomer come out
naturally from the underlying Lagrangian.

\section{nuclear equation-of-state}
Our study is based on the
relativistic transport model for the expansional stage of the fireball
formed in a Si+Pb collision at 14.6 (GeV/c)/nucleon \cite{KOCH91,FANK93}.
In Ref. \cite{LI94A}, we have proposed two sets of parameters for
the non-linear
$\sigma$-$\omega$ model. In this work, we use slightly different parameters
which are determined by requiring a saturation density of 0.16 fm$^{-3}$,
a binding energy of 15.96 MeV, a nucleon effective mass $m^*/m$=0.77, and an
incompressibility of 380 MeV.
Using the notation of Ref. \cite{LI94A}, these parameters have the following
values
$$C_V= 10.5, ~~C_S=13.61, ~~B=-0.003131, ~~C=0.02908.$$
The nuclear
equation-of-state corresponding to this parameter
set is shown in Fig. 1 by the dashed curve. To simulate the
density isomer at high densities, we assume that the vector coupling
constant is density-dependent and
is determined by nuclear matter properties at saturation density
and the assumption that around 5.5$\rho _0$ a secondary minimum appears.
Explicitly, we have
\begin{eqnarray}
C_V&=&10.5, ~~\rho \le \rho _0,\nonumber\\
&=&10.59-0.0923\Big({\rho \over \rho_0}\Big)^2, ~\rho _0<\rho \le 5.5
\rho _0,\nonumber\\
&=&7.8, ~~\rho > 5.5\rho _0.
\end{eqnarray}
The nuclear equation-of-state with this density-dependent
vector coupling constant is plotted in Fig. 1 by the solid curve.
A secondary
minimum is seen around 5.5$\rho _0$ with a binding energy of about 2 MeV.

We note that the density dependence of the vector coupling constant leads to
a so-called rearrangement potential for the nucleon. In the present case,
this potential is attractive, so it tends to further slow down the
expansion of the fireball.

The fireball, which contains nucleons ($N$), deltas ($\Delta$), hyperons
($Y$), pions ($\pi$), rhos ($\rho$), as well as kaons ($K^+$) and
antikaons ($K^-$), is assumed to have an initial density of 6$\rho _0$ and
temperature of 170 MeV. The initial density is higher than that used
in the fireball model of Ref. \cite{KO91} but is consistent with
that predicted by the
cascade model \cite{PANG92} and the relativistic quantum molecular dynamics
\cite{MATT89}.
The motions of these particles are then described by
the relativistic transport model.  The treatment of two-body scattering
is the same as in Refs. \cite{FANK93,LI94A}. For kaon-baryon and
antikaon-baryon elastic scattering we use the empirical
cross sections \cite{DOV82}. The kaon-pion and anitkaon-pion
scattering are dominated by $K^*$ resonances, and we use the
Breit-Wigner cross section of Ref. \cite{KO81}.
For an antikaon, it can also be annihilated by a nucleon into a hyperon
and a pion ($K^-N\rightarrow Y\pi $).  The annihilation cross section
is appreciable and is expected to be modified strongly in a medium due to
changing in-medium hadron masses.  The in-medium cross section
for antikaon annihilation will be discussed later.

\section{kaon and antikaon potentials}
In a nuclear medium, a kaon feels an attractive scalar potential due to
explicit chiral symmetry breaking and a repulsive vector potential
due to $\omega$-exchange. The vector potential for an antikaon is attractive
under the G-parity transformation. In mean-field approximation,
the kaon and antikaon dispersion relations can be
written as
\begin{eqnarray}
\omega ^{*2}&=& m_K^2+{\bf k}^2_K-{\Sigma _{KN}\over f_K^2}\rho _S
\pm {3\over 4}{\omega_K^*\over f_K^2}\rho _B\nonumber\\
&=&m_K^2+{\bf k}^2_K-{\Sigma _{KN}\over f_K^2}\rho _S
\pm {2\over 3}\omega_K^*\Big({g_\omega\over m_\omega}\Big)^2\rho _B,
\end{eqnarray}
where $\rho _B$ and $\rho _S$ are the baryon and scalar densities,
respectively.
The second line follows from the
SU(3) relation $g_\omega =3g_\rho $ and the KFSR relation $m_\rho
=2\sqrt 2 f g_\rho$ (assumption of $f_K=f_\pi =f$ is implied) \cite{BROW92}.

In Eq. (2) the plus and minus signs are for kaon and antikaon,
respectively, $\Sigma _{KN}$ is the KN sigma term, and $f_K$ is the
kaon decay constant. In this work, we use $\Sigma _{KN}= 350$ MeV and
$f_K$= 93 MeV, as used in our study of
kaon and antikaon production in heavy-ion collisions at SIS
energies \cite{FANG94,LI94B}.
Following Ref. \cite{SHU92}, we define the kaon (antikaon) potential
as
\begin{eqnarray}
U_{K,{\bar K}}(\rho ,{\bf k})=\omega ^*(\rho ,{\bf k})-\omega _0({\bf k}).
\end{eqnarray}
The results for the kaon (antikaon) potential with zero momentum in
a nuclear medium is shown in Fig. 1. One sees that
the kaon potential is weakly repulsive at low densities,
qualitatively consistent with that from the impulse approximation
based on the kaon-nucleon ($KN$) scattering length. At high densities (above
about 3$\rho _0$) the kaon potential becomes attractive as a result
of the reduced
repulsive vector interaction. The antikaon potential is attractive
at all densities.

The propagation of kaons and antikaons in their mean-field potentials
is given by the following equations of motion
\begin{eqnarray}
{d{\bf x}\over dt} ~= ~ {{\bf k}\over E^*}, ~~~{d{\bf k}\over dt}
{}~=~ - \nabla _xU_{K,{\bar K}} (\rho ,{\bf k}),
\end{eqnarray}
where $E^*=\Big[m_K^2+{\bf k}^2-(\Sigma _{KN}/f_K^2)\rho_S+
(\frac{1}{3}\frac{g_\omega^2}{m_\omega^2}\rho_B)^2\Big]^{1/2}$,
and $U_{K,{\bar K}}$ is given by Eq. (3).
When a kaon (antikaon) propagates from the high density to the
low density (especially when it crosses the surface of the fireball),
its momentum is reduced by the attractive potential. This
is the mechanism by which the attractive potential cools down particles.

\section{antikaon annihilation in medium}
A main difference between a kaon and an antikaon in a
dense matter is that the
latter can be annihilated by a nucleon into a hyperon and a pion, i.e.,
$K^-N\rightarrow Y\pi $. This cross section at low energies has been
analysed by the K-matrix method of Martin {\it et al.} \cite{MAR81}, and
the results can be fitted by the following
parameterization
\begin{eqnarray}
\sigma _{ann.} (\sqrt s) ~= ~ {134.8\over 1.0+76.7 (\sqrt s- \sqrt {s_0})
+ 161.1 (\sqrt s-\sqrt {s_0})^2},
\end{eqnarray}
where $\sqrt s$ is the invariant energy of the $K^-N$ system and $\sqrt {s_0}
=m_N+m_K$. In nuclear medium, we assume that the form of the annihilation
cross section does not change, but replace $\sqrt s$ and $\sqrt {s_0}$ by
$\sqrt {s^*}$ and
$\sqrt {s_0^*}=m^*_{\bar K}+m^*_N$, respectively.
Another important medium effect is that the reaction $K^-N\rightarrow Y\pi$
may be forbidden in the medium.  To see this, we take
the scalar potential of an antikaon to be
approximately 1/3 of the nucleon scalar potential $U_S$ and that of hyperon
to be $(2/3)U_S$.  In the reaction $K^-N\rightarrow Y\pi$,
the total scalar potential is then (4/3)$U_S$ in the initial state
and is more negative than that in the final state, which is $(2/3)U_S$.
For antikaons with
low momenta, which we are interested in, there is thus a possibility that
the invariant energy $\sqrt {s^*}$ of the antikaon and nucleon
is below $m_Y^*+m_\pi$, so
the antikaon cannot be annihilated.

\section{results}

Because of the density isomer around $5.5\rho _0$, the fireball expands
rather slowly. In Fig. 3, we show the baryon density profile of the fireball
at different times. In the first 4 fm/c, the central density decreases by only
about 10\%. From t=4 fm/c to 8 fm/c, the central density
decreases from 5.5$\rho _0$ to about 2.0$\rho _0$. Up to 10 fm/c, a
well-defined surface is seen, and when kaons move across this surface their
momenta are changed.

To obtain good statistics, we use the perturbative
test particle method of Ref. \cite{FANG93} for kaons and antikaons.
Furthermore, we average the transverse mass spectra of kaons
and antikaons over the entire rapidity range as in Ref.
\cite{KOCH94}. Using a bin size of 5 MeV, we have more than one thousand
events within each transverse mass bin. In Fig. 4, we show the transverse
mass spectra of kaons at different times. The upper-left panel gives the
initial kaon spectrum which are in an equilibrium fireball at temperature
$T=170$ MeV. The spectrum can be fitted by ${\rm exp}\big(-m_t/T^\prime\big)$
with an apparent temperature $T^\prime\sim 145 $MeV, which is smaller
than the source temperature $T$. This difference is mainly due to the
rapidity effect, as the apparent temperature $T^\prime$ is related to the
source temperature $T$ by $T=T^\prime$cosh ($y$), where $y$ is the
kaon rapidity in the center-of-mass frame of the fireball.

The upper-right panel gives the kaon spectrum after 4 fm/c. During
this time interval, kaons with high momenta have mostly escaped the potential
well. We already see an enhancement of low-$m_t$ kaons and  a weak
low temperature component. The steep rise of the extremely low-$m_t$ kaons
happens mainly during the time interval from t=4 fm/c to 8 fm/c, as
shown in the lower-left panel of the figure. During this time interval,
kaons that are initially trapped inside the potential well gradually
climb out the well as the system slowly expands. A well-defined
low temperature component is visible at this time.
Afterwards there is little change in the
kaon spectrum, as shown in the low-right panel for t=12 fm/c,
as the density of the system is already below 2$\rho _0$ and
the kaon potential is very weak.

In Fig. 5, we compare our results with preliminary experimental data
of the E814 collaboration \cite{STA94}. The experimental data are in the
forward rapidity ($2.2\ge y <2.4$), while our results are averaged over
the entire rapidity region as mentioned early. Since we do not
have a complete dynamical model for the entire process of heavy-ion
collisions at AGS energies, we scale our final spectrum such that
we fit the data point at $m_t-m_K \sim 7 $ MeV.
The initial spectrum, given by the dotted histogram,
is similarly scaled, and can be characterized by one
apparent temperature of about 145 MeV.
The final kaon spectrum, obtained after including both
propagation in the mean-field potential
and rescattering with baryons and pions,
is given by the solid histogram and shows clearly a
two-component structure.
The component corresponding to high transverse masses ($m_t-m_K\ge
0.02$  GeV) has an apparent temperature of about 140 MeV,
and is similar to the initial kaon spectrum.
The steep rise of low-$m_t$ kaons ($m_t-m_K<0.02$ GeV)
can be fitted by an exponential with an apparent temperature of
about 40 MeV. Although this temperature is still about
a factor of two larger than  the temperature
extracted from the experimental data,
it is considerably smaller than that of kaons with high
transverse masses.

Finally, we compare in Fig. 6 our results for antikaon spectra with
preliminary data of Ref. \cite{STA94}. Again,  we
scale our final antikaon spectrum such that we fit the data point at
$m_t-m_K\sim 7 $ MeV.
Because of annihilation, the final antikaon
multiplicity is smaller than the initial one. For comparison,
we have reduced the initial spectrum by
the ratio of the final antikaon multiplicity to the initial one.
The initial spectrum (dotted histogram) can be fitted by a single exponential
with an apparent temperature of about 145 MeV.  The final antikaon
spectrum again shows a two-component structure. The high-$m_t$
component is similar to the initial spectrum, while the low-$m_t$
component has a much lower temperature of about 30 MeV, which is again
about a factor of two larger than that extracted from the experimental data.
The $K^-$ temperature is lower than that of $K^+$ as its potential remains
attractive throughout the whole expansion of the fireball while the kaon
potential becomes repulsive once the density is below about 3$\rho_0$.
We note that the reduced annihilation of low-energy $K^-$ in dense matter as
discussed in section IV is crucial in obtaining the low-$m_t$ antikaons.
Finally, we mention that in our calculation the proton spectrum does not show
a two-component structure, which is consistent with the experimental data.

\section{conclusions}
In summary, we have studied the enhancement of low-$m_t$ kaons and
antikaons in heavy-ion collisions at AGS energies
in the relativistic transport model.
By introducing a density isomer
at high densities, the expansion of the fireball is slowed down
so that kaons and antikaons can be effectively cooled
down by their attractive mean-field potentials. We see clearly
a `cold' low-$m_t$ component in both kaon and antikaon transverse mass
spectra, as  observed in AGS experiments by the E814 collaboration.
To carry out more detail comparisons with the experimental measurements,
we need to introduce rapidity cuts in analyzing the kaon and antikaon
spectra.  Also, it is necessary to extend the relativistic transport
model to include more degrees of freedom so that it will be suitable
for describing the entire collision dynamics at AGS energies.

\vskip 1cm

\noindent{\bf ACKNOWLEDGEMENT}
\bigskip

We are grateful to Gerry Brown and Volker Koch for helpful conversations.
This work was supported in part by the National Science Foundation under
Grant No. PHY-9212209 and the Welch Foundation under Grant No. A-1110.
We also thank the [Department of Energy's] Institute for Nuclear Theory
at the University of Washington for its hospitality and the Department
of Energy for partial support during the completion of this work.

\pagebreak

{\bf Figure captions}:

{\bf Fig. 1}: ~Equation-of-state of nuclear matter. The dashed curve
is based on the non-linear $\sigma$-$\omega$ model parameters of Ref.
\cite{LI94A}. The solid curve is based on the density-dependent vector
coupling constant of Eq. (1).

{\bf Fig. 2}: ~Kaon and antikaon potentials (with zero momentum) in
nuclear matter as a function of baryon density.

{\bf Fig. 3}: ~Baryon density profile of the fireball at different times.

{\bf Fig. 4}: ~Kaon transverse mass spectra at different times.

{\bf Fig. 5}: ~Initial and final kaon transverse spectra. The experimental
data for $2.2\ge y<2.4$ are from Ref. \cite{STA94}.

{\bf Fig. 6}: ~Same as Fig. 4 for antikaons.

\end{document}